# Synthetic CT Generation via Variant Invertible Network for All-digital Brain PET Attenuation Correction

Yu Guan, Bohui Shen, Xinchong Shi, Xiangsong Zhang, Bingxuan Li*, Qiegen Liu*, *Senior Member, IEEE*

*Abstract*——Attenuation correction (AC) is essential for the generation of artifact-free and quantitatively accurate positron emission tomography (PET) images. However, AC of PET faces challenges including inter-scan motion and erroneous transformation of structural voxel-intensities to PET attenuation-correction factors. Nowadays, the problem of AC for quantitative PET have been solved to a large extent after the commercial availability of devices combining PET with computed tomography (CT). Meanwhile, considering the feasibility of a deep learning approach for PET AC without anatomical imaging, this paper develops a PET AC method, which uses deep learning to generate continuously valued CT images from non-attenuation corrected PET images for AC on brain PET imaging. Specifically, an invertible network combined with the variable augmentation strategy that can achieve the bidirectional inference processes is proposed for synthetic CT generation (IVNAC). To evaluate the performance of the proposed algorithm, we conducted a comprehensive study on a total of 1440 data from 37 clinical patients using comparative algorithms (such as Cycle-GAN and Pix2pix). Perceptual analysis and quantitative evaluations illustrate that the invertible network for PET AC outperforms other existing AC models, which demonstrates the potential of the proposed method and the feasibility of achieving brain PET AC without CT.

## I. Introduction

Positron-emission tomography (PET) as a non-invasive functional imaging modality, provides three-dimensional distribution maps of radioactive tracers at the molecular and cellular level enabling the in-vivo estimation of chemical and biological processes of the tissues [1]. Therefore, it plays an indispensable role in early disease detection and postoperative patient staging diagnosis. Generally, attenuation correction (AC) of annihilation photons is a critical procedure in PET image generation for providing accurate quantitative information on the radiotracer distribution. However, the main challenge of AC lies in finding reliable attenuation-correction factors (ACF) compensating for this loss before or during image reconstruction.

Traditionally, the ACF is measured in an additional PET transmission scan. However, the transmission scan adds at least 50% to the scanning time and results in data with a relatively high noise level [2]. When these data are used for corrections, they propagate the noise into the emission images. In a combined PET/CT system, due to the ability of CT imaging to reveal accurate anatomical information with good contrast of soft tissue and bone, it is used to generate the ACF by simple piecewise scaling of a CT image [3]. However, deriving accurate ACF from additional anatomic imaging (e.g., the X-rays or gamma rays during CT imaging) exposes the patients to extra doses of ionizing radiation. To reduce the dose of ionizing radiation exposure as much as possible while simultaneously obtaining precise anatomic imaging for PET AC, various methods have been developed to replace CT imaging with other manners.

On the one hand, PET AC can be assisted with other non-radiation devices. Among them, the incorporation of magnetic resonance (MR) imaging with PET (PET/MR) [36-38] becomes a promising alternative to the existing PET/CT system by providing excellent soft tissue visualization without ionizing radiation. Nevertheless, estimation of the required ACF is based on MR images and is particularly challenging because the bone and tissue with the largest attenuation coefficient, are not visible with positive contrast under typical MR acquisitions [4]. Additionally, even with advanced acquisitions, bone structure and air often remain difficult to distinguish. Research interest in alternative approaches (such as LU177-based detectors) for PET AC has grown substantially in the last years. The estimation of $\mu$ maps can be accomplished by harnessing the background intrinsic activity of Lu177-based detectors, which utilize crystals with large axial field of view (FOV) in PET systems [39], [40]. However, the limitation in this case is that they rely on a specific class of detectors and relatively long acquisition times to obtain adequate background inherent count statistics. Another emerging technology, time-of-flight (TOF) PET scanners, could potentially offer an elegant solution to the AC problem. Note that maximum-likelihood activity and attenuation (MLAA) is one of the techniques designed for generating attenuation maps in TOF PET systems without anatomical imaging. The technique iteratively estimates activity and attenuation information by solving a joint estimation problem. Though MLAA has already been shown that can produce improved reconstruction results in TOF cases, it is still unclear whether the uniqueness also applies to the Poisson likelihood, except if the TOF data are consistent. On the other hand, a potential strategy for PET AC is to directly generate synthetic CT images from another modality images, thus avoiding the problem of additional acquisition of CT images. Obviously, this idea effectively avoids the discomfort of patients caused by multiple scans

This work was supported in part by the National Natural Science Foundation of China under 61871206, 61601450 and 62201193. (Corresponding authors: Bingxuan Li; Qiegen Liu.)

This work did not involve human subjects or animals in its research.

Y. Guan, B. Shen and Q. Liu are with the Department of Electronic Information Engineering, Nanchang University, Nanchang 330031, China. ({guanyu, shenbohui}@email.ncu.edu.cn, {liuqiegen}@ncu.edu.cn).

X. Shi and X. Zhang are with the Department of Nuclear, First affiliated hospital of Sun Yat-Sen University, Guangzhou 510000, China. ({shixch, zhxiangs}@mail.sysu.edu.cn).

B. Li is with the Institute of Artificial Intelligence, Hefei Comprehensive National Science Center, Hefei 230088, China. (libingxuan@iai.ustc.edu.cn).

and removes extra CT radiation. Thus, PET-only AC methods are still urgently needed.

Inspired by the rapid expansion of deep learning in both industry and academia in recent years, many research groups have attempted to integrate deep learning-based methods into medical imaging and radiation therapy [5]. Therefore, direct application of AC in the image domain using deep learning approaches has been proposed for the dedicated PET systems that lack accompanying transmission or anatomical imaging [6], [7]. It is further considered that CT is able to provide the ACF to correct annihilation photon losses due to attenuation processes in the patient body. However, CT is unavailable in PET-only scanners or PET/MR scanners which do not provide photon attenuation coefficients. Current studies proposed several schemes to address this issue, a conceptually-straightforward approach is to use a deep network to generate synthetic CT images from non-attenuation-corrected PET (NAC PET) images for AC on PET imaging [8]. For example, an automated PET image AC approach was proposed for the first time [9] which synthetic CT images were estimated from NAC PET images with an encoder-decoder network to perform AC for quantitative PET imaging. Hashimoto et al. [10] developed a deep-learning-based PET AC framework that synthesizes transmission CT images from NAC PET images using a convolutional neural network for brain-dedicated PET scanners. Interestingly, Guo et al. [11] simplified the complex problem by a domain decomposition for PET attenuation so that the learning of anatomy-dependent AC can be achieved robustly in a low-frequency domain while the original anatomy independent high-frequency texture can be preserved during the processing. Going in a similar direction, PET AC is also conducted using generative adversarial networks (GANs) to synthesize CT images and to directly generate AC PET images. A deep learning algorithm based on GAN was implemented and trained to generate synthetic CT data from NAC PET data [12]. Moreover, Dong et al. [13] proposed a method that structures a cycle generative adversarial network (Cycle-GAN) to establish the transformation that minimizes the difference between the synthetic CT generated from NAC PET and the true CT. However, these approaches are vulnerable to outliers and fail to recover quantitative accurate activity around the center of the head with complex anatomical structures.

In another application domain, there is an alternative approach to directly learn the relationship mapping between NAC PET and AC PET images, i.e., NAC-AC PET learning. Hu et al. [14] presented a deep-learning-based approach to estimate synthetic AC PET images and synthetic CT images for whole-body PET imaging by taking full advantage of existing NAC PET images. Another strategy that focused on estimating synthetic AC PET images from NAC PET images directly using Cycle-GAN was also proposed [15]. Subsequently, a pix2pix model was adopted to directly predict AC PET images from NAC PET images by Li et al., which allowed the calculation of PET attenuation maps [16]. However, these proposed frameworks require the reconstruction of NAC PET images first, which may increase computational time compared to the above-mentioned method of generating synthetic CT images from NAC PET images. Accordingly, it is of interest whether other deep learning approaches can mitigate the limitations of independent PET AC which does not require any other imaging modality and is thus only based on PET data.

The purpose of this study is thus to develop and implement an invertible network-based deep learning method (IVNAC) that allows the generation of synthetic CT using only a brain NAC PET image and also reduces the imaging computation time. To generate CT images without the use of additional acquisition of either MR images or transmission images, the variable augmentation ideology is incorporated into the invertible network to enhance the relevance of data, which is conducive to the generation of characterization information from NAC PET images. Meanwhile, we proposed the concept of all-digital PET for the first time and has already been commercialized [41], [42]. All-digital PET can obtain accurate scintillation pulse waveform data, which has the characteristics of precise sampling, so the datasets obtained by all-digital PET equipment has higher resolution [43-45]. Moreover, the architecture of all-digital PET offers the opportunity to integrate new digital signal processing principles in the medical imaging field, which is characterized by excellent accuracy, modularity, flexibility, and self-adjusting systems, will provide unprecedented performance and reliability for molecular imaging research applications.

The main contributions of this work are summarized as follows:
- To the best of our knowledge, this is the first work that introduces the invertible network to tackle the problems of PET AC in a unified framework. NAC PET images share similar anatomical structures to the AC PET images, but the lack of detailed information makes it difficult for end-to-end mapping reconstruction. Therefore, synthetic CT images are introduced as an intermediate for PET AC. Meanwhile, to learn the difference between NAC PET and synthetic CT images, variable augmentation ideology is adopted to enhance the generative power of the invertible network.
- All-digital PET is previously shown to significantly improve the quality of PET images due to an enhanced time-of-flight resolution and to better prevent the count-related rises in dead time and pile-up effects due to the use of digital systems with small trigger domains. Meanwhile, the availability of a high-quality dataset serves as a fundamental basis for enhancing model performance and establishing superior prerequisites for subsequent PET AC.

## II. PRELIMINARIES

### A. Direct Prediction of Attenuation Corrected PET

It is widely acknowledged that CT serves as the gold standard approach for PET AC. CT exhibits excellent electron density contrast capabilities, enabling precise AC in PET imaging. In contrast, MRI offers superior soft tissue contrast compared to CT while its limitations in quantifying electron density prevent its standalone use for PET AC. Therefore, a conceptually-straightforward approach for PET AC is to transform CT-derived or MR-derived synthetic CT images into attenuation maps [17], [18].

According to recent studies of the remarkable performance of deep learning and various PET AC algorithms [14-16], it is currently common to use deep networks to predict attenuation-corrected PET images from NAC PET images. In this paradigm, NAC PET images are regarded as the input to a deep learning network to generate the synthetic CT images and then predict the attenuation corrected PET images. Notably, the synthetic CT values expressed in HU were converted to linear attenuation coefficients in $cm^{-1}$ by a piecewise linear transformation to obtain the attenuation map ($\mu$-map) [19].

$$\mu = \begin{cases} \mu_{water}(1 + \dfrac{I^{CT}}{1000}) & I^{CT} \leq 0HU \\ \mu_{water}(1 + \dfrac{I^{CT}}{1000}\dfrac{\rho_{water}(\mu_{bone} - \mu_{water})}{\mu_{bone}(\rho_{bone} - \rho_{water})}) & I^{CT} > 0HU \end{cases} \quad (1)$$

where $\mu_{water}$ and $\mu_{bone}$ represent the attenuation coefficients at the PET 511 keV energy for water and bone, respectively. $I^{CT}$ represent CT values. $\rho_{water}$ and $\rho_{bone}$ represent the attenuation coefficients at the CT energy (120 keV), respectively. All parameter values are described as follows: $\mu_{bone} = 0.172 cm^{-1}$, $\mu_{water} = 0.096 cm^{-1}$, $\rho_{bone} = 0.326 cm^{-1}$, $\rho_{water} = 0.158 cm^{-1}$.

### B. Basic Theory of Invertible Network

Generally, multiple layers of neurons and weighted connections between the layers are included in traditional neural networks [20]. Although activation functions of the neurons in the hidden layers enable the neural networks to capture the nonlinear relation between the input and output data, they only focus on forward prediction processes. Therefore, they do not have inverse inference capability, due to the limitation of the network structure. To handle the inverse process, invertible network models have been proposed [21], [22].

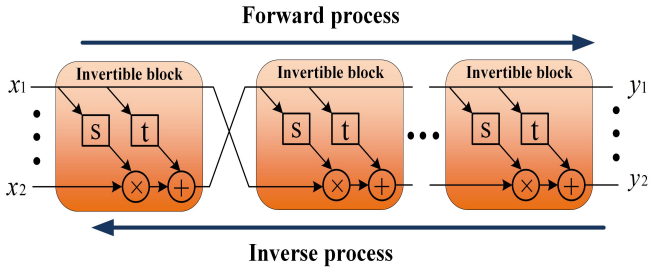

**Fig. 1.** Structure of invertible network consisting of invertible blocks.

The invertible network consists of 8 invertible blocks, where each invertible block includes a 1x1 invertible convolution and affine coupling layers [23], [24]. The general architecture of the invertible network is shown in Fig. 1. Notably, each invertible block needs an even and equal number of inputs and outputs, which consists in splitting the input $x$ into two sub-vectors $x_1$ and $x_2$ of equal size, then successively applying two (no necessarily invertible) functions following the equation:

$$\begin{cases} y_1 = x_1 \\ y_2 = x_2 \odot \exp(s(x_1)) + t(x_1) \end{cases} \quad (2)$$

where $\odot$ represents the element-wise multiplication, $y_1$ and $y_2$ correspond to the two parts that compose the output vector $y$, $s$ and $t$ are arbitrary nonlinear functions which are not necessarily invertible. Particularly, the functions $s$ and $t$ can also represent neural networks themselves, such as densely connected convolutional neural networks [33], which can be learned during training. Based on the above formulation, the inverse process can be computed as:

$$\begin{cases} x_1 = y_1 \\ x_2 = (y_2 - t(y_1)) \odot \exp(-s(y_1)) \end{cases} \quad (3)$$

It can be seen that the invertible network learns the mapping as $y = f(x)$, which is fully invertible as $x = f^{-1}(y)$. Hence, computing the inverse of the block does not bring additional computational complexity compared to the forward process. Furthermore, because of the unique bijective property of the model structure, it can achieve the invertibility of one single network and allow for bidirectional operation and training. Moreover, the information is fully preserved during both the forward and reverse transformations.

## III. METHODOLOGY

### A. Generation of Synthetic CT Images for PET-only AC

As mentioned above, the invertible network is selected as the footstone of the proposed AC method in this work due to the properties such as reversibility, memory savings, and simple design. More specifically, a deep learning algorithm based on the invertible network is implemented and trained to generate synthetic CT data from NAC PET data for subsequent brain PET AC. The proposed processing pipeline consists of two independent phases for training retrospective data and reconstructing new data, respectively. As shown in Fig. 2, the proposed invertible network consists of 8 invertible blocks, and the training data is composed of NAC PET images and reference CT images as inputs. More precisely, the reference CT images are used as learning targets for the corresponding NAC PET images in the basic process of generating CT from NAC PET. Subsequently, the invertible network iteratively estimated synthetic CT images and compared them with the reference CT data.

During the training stage, we compute the deviation of the model output $y_{CT}^{i-1}$ from the network prediction $f(x_{PET})$ with a loss denoted $L_{y_{CT}}(y_{CT}, f(x_{PET}))$, where $L_{y_{CT}}$ is the **loss1** associated with the label and the invertible model. In case of the forward process, the goal of the network is to map a NAC PET image $x_{PET}$ to the corresponding reference CT image $y_{CT}^{i-1}$ through the mapping function. This can generally be viewed as a regression task between two image domains that share the same underlying structures but differ in surface appearance. Subsequently, prior information is extracted from the $x_{PET}$ and fed into the invertible network, which maps NAC PET patches to synthetic CT images. Due to invertibility, the corresponding inverse process of the model is implicitly learned through **loss2**. In detail, the inverse pass of the invertible network provides more complete prior information by inversely mapping $y_{CT}^{i-1}$ to $x_{PET}^{i-1}$. Since the information lossless property of the invertible network can preserve the detailed information of the input data, and the constant mapping can solve the problem of network degradation and ensure the stability of its data generation, the invertible network has powerful advantages in the generation of synthetic CT image [34], [35].

### B. Invertible and Variable Augmented Network

To better improve the learning performance of the invertible network for synthetic CT image generation, we add dummy variables and replicate them in the network based on the concept of variable augmentation [25]. More specifically, a new high-dimensional invertible network architecture different from the traditional invertible network structure is proposed, which not only has strong robustness and fault tolerance but also improves the training and learning ability by the variable augmentation technology. High-frequency information often includes irrelevant and uninformative components since NAC PET images contain a large amount of noise. These components may prevent the model from obtaining an accurate mapping from NAC PET to CT. The introduction of variable augmentation strategy overcomes the difficulty, it can handle both quantitative and qualitative information, and can well coordinate multiple input information relationships to identify the most informative components and mitigate noise interference, so it is suitable for generating high-quality CT images.

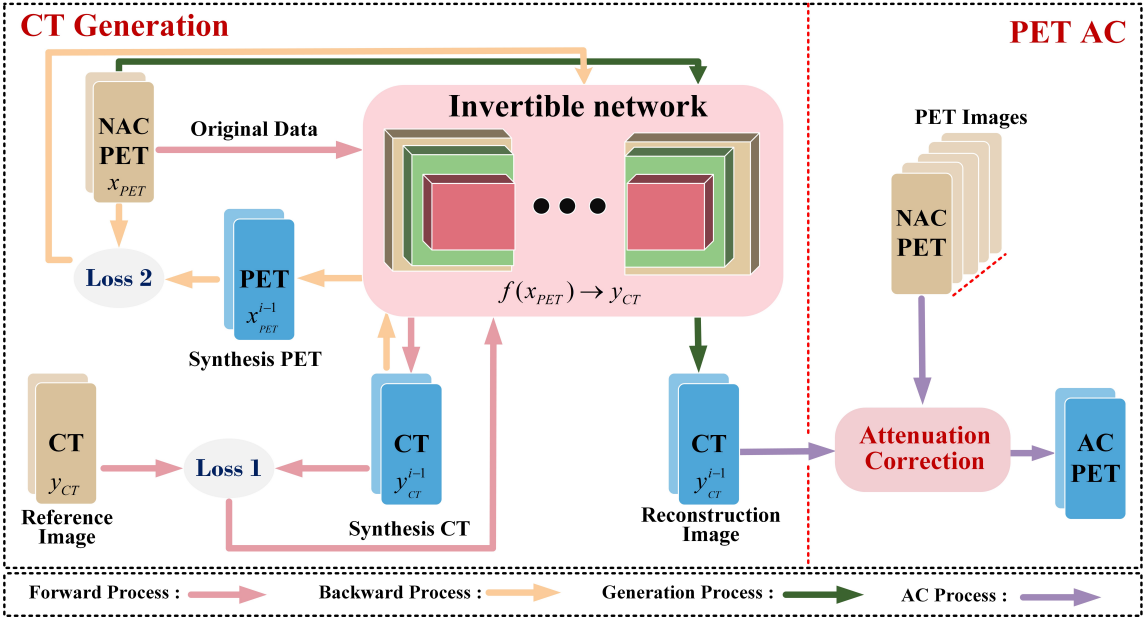

Fig. 2. The schematic flow diagram of the proposed method. The training phase is first performed with NAC PET and reference CT images, after which the well-trained network is fixed and ready for generating synthetic CT images for new PET data in the reconstruction phase.

In addition to improvements to the invertible network, it is also necessary to find a bijective function that can map the data point from NAC PET image $x_{PET}$ to generated CT image $y_{CT}$ and realize the process of reversibility simultaneously. There are various ways of implementing bijection functions. Classical methods need two separate networks to approximate $y = f(x)$ and $x = f^{-1}(y)$ mappings respectively, which leads to inaccurate bijective mapping and may accumulate the error of one mapping into the other. Following the guidelines to enable the invertibility of one single network, we take an alternative method and use the affine coupling layers described in [23], [26]. Subsequently, by combining a series of invertible and tractable bijective functions $\{f_i\}_{i=0}^{I}$, we construct the bijective function of the invertible and variable augmentation network, i.e., $f = f_0 \circ f_1 \circ f_2 \circ \cdots \circ f_I$ ($\circ$ represents the multiplication of a map, i.e., a composite map). For a given observed NAC PET image $x_{PET}$, we can derive the transformation to synthetic CT image $y_{CT}$ through the following formula:

$$y_{CT} = f_0 \circ f_1 \circ f_2 \circ \cdots \circ f_I(x_{PET}) \quad (4)$$
$$x_{PET} = f_I^{-1} \circ f_{I-1}^{-1} \circ \cdots \circ f_0^{-1}(y_{CT}) \quad (5)$$

The objective model $f_i(\cdot)$ is implemented through affine coupling layers. An affine coupling layer performs a reversible nonlinear transformation, which indicates that it can learn not only the inverse mapping but also the parameter mapping at no additional computational cost. In each affine coupling layer, the input and output are denoted as $x_{PET}$ and $y_{CT}$, respectively.

Due to the need of each invertible block for an even and equal number of inputs and outputs, the input is divided into two halves $x_{PET} = (x_{PET}^1, x_{PET}^2)$, which are transformed by an affine function:

$$y_{CT}^1 = x_{PET}^1 \odot \exp(s_1(x_{PET}^2)) + t_1(x_{PET}^2) \quad (6)$$
$$y_{CT}^2 = x_{PET}^2 \odot \exp(s_2(y_{CT}^1)) + t_2(y_{CT}^1) \quad (7)$$

Since $s$ and $t$ are arbitrary nonlinear functions, here $s_{(\cdot)}$ and $t_{(\cdot)}$ represent scale and translation functions, respectively. Subsequently, the outputs $y_{CT} = (y_{CT}^1, y_{CT}^2)$ are connected again and passed to the next affine coupling layer. The inverse step is easily obtained as follows:

$$x_{PET}^2 = (y_{CT}^2 - t_2(y_{CT}^1)) \odot \exp(-s_2(y_{CT}^1)) \quad (8)$$
$$x_{PET}^1 = (y_{CT}^1 - t_1(x_{PET}^2)) \odot \exp(-s_1(x_{PET}^2)) \quad (9)$$

Next, we utilize the invertible 1×1 convolution proposed in [23] as the learnable permutation function to reverse the order of channels for the next affine coupling layer.

In summary, the unique structure of bijective function permits bidirectional training of the network. Meanwhile, combining the variable augmentation strategy and the invertible nature of the bijective function, the effective information between data generation is preserved. The general IVNAC architecture is illustrated in Fig. 3. The input image is split into two halves along the channel dimension. $s$, $t$ and $r$ are transformation equal to dense block, which consists of five 2D convolution layers with filter size $3 \times 3$. Each layer learns a new set of feature maps from the previous layer. The size of the receptive field for the first four convolutional layers is $3 \times 3$, and stride is 2, followed by a rectified linear unit. The last layer is a convolution without ReLU. The purpose of the Leaky ReLU layers is to avoid overfitting to the training set [27] and further increase nonlinearity, which can improve the learning ability to generate higher resolution CT images. In the forward process, the input image $x_{PET}$ is transformed to output images of other modalities by a stack of bijective functions $\{f_i\}_{i=0}^{I}$.

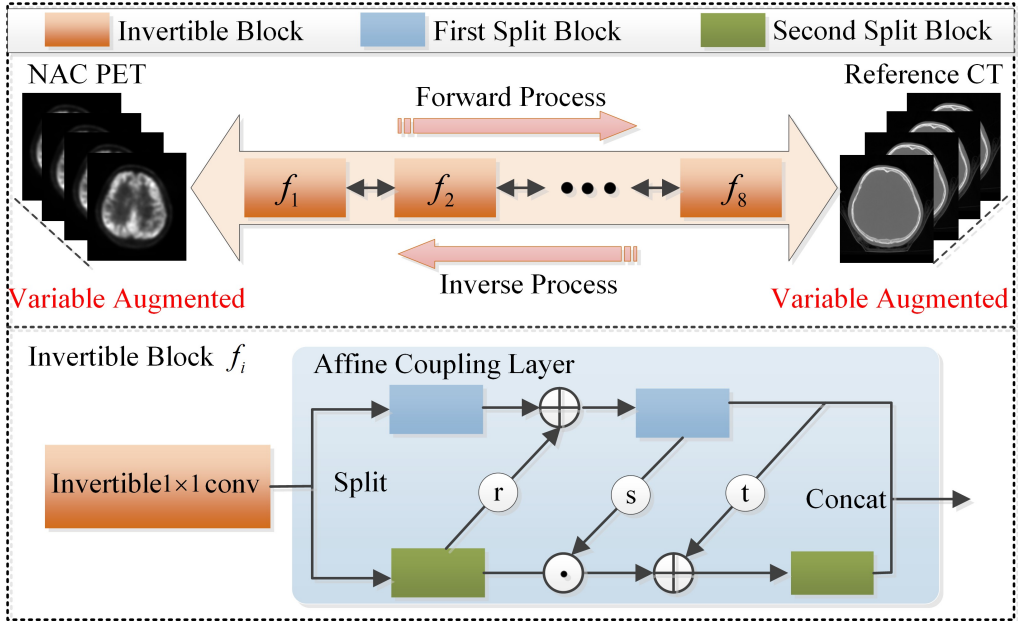

**Fig. 3.** The pipeline of IVNAC. Invertible model is composed of both forward and inverse process. We illustrate the details of the invertible block on the bottom. $s$, $t$ and $r$ are transformations defined in the bijective functions $\{f_i\}_{i=0}^{I}$.

*C. Multi-components Loss Function*

As described above, the network relies on continuous improvement of a sequence of invertible blocks including affine coupling layers, actnorm layers and so on. The accuracy of both invertible blocks is directly dependent on the design of the corresponding loss function. Hence, multi-component loss function is leveraged to optimize the network to guarantee the quality of the generated images. For example, in MM-Synthesis [28], a multi-input multi-output fully convolutional network model for PET synthesis is trained by a cost function constituted from three cost components. In MM-GAN [29], generation loss and adversarial loss constitute the cost function of this network. The Hi-Net [30] network adds a reconstruction loss on the basis of the generation and discriminator loss. Similarly, not only the FGEAN [31] includes pixel-wise intensity loss, it also includes gradient information loss. Both of them use multiple loss functions to make sure that the network architecture performs well in a variety of generation tasks.

Invertible network doubles the process of training by enforcing an inverse mapping constraint on the model, improving the accuracy of synthetic CT images. Therefore, based on the unique invertible nature of the network, multi-component loss function is used to guide the invertible network to simultaneously capture both the high and low-frequency components of the NAC PET images in both directions, thereby achieving precise multi-modal image transformation (generated from NAC PET images to synthetic CT images). Furthermore, compared with the above-mentioned methods, our multi-component loss function based on Euclidean loss has advantages of simple composition and easy training.

Schematic flowchart of the training process for the multi-component loss function is outlined in Fig. 4. It minimizes the mean squared error between pixel values of the NAC PET image and the reference CT image, for clarity, it can be mathematically expressed as follows:

$$\mathcal{L}_{total} = \lambda\mathcal{L}_1 + \mathcal{L}_2 = \lambda\|f(x_{PET}) - y_{CT}\|_2 + \|f^{-1}(y_{CT}) - x_{PET}\|_2 \quad (10)$$

where $y_{CT}$ is the reference CT image, $f(x_{PET})$ is the output image from the source image $x_{PET}$ by invertible network $f(\bullet)$, and $\|\cdot\|_2$ represents the $\mathcal{L}_2$-norm. $\mathcal{L}_1$ stands for the loss function between the synthetic CT image and reference CT image. $\mathcal{L}_2$ stands for the loss function between the inverted output image and NAC PET image. $\lambda$ is a hyperparameter used to balance the two losses. Notably, the implicit variables $z$ in Fig. 4 are used as additional latent output variables to capture the information otherwise lost, that is, to store information missing from the forward mapping from $x$ to $y$.

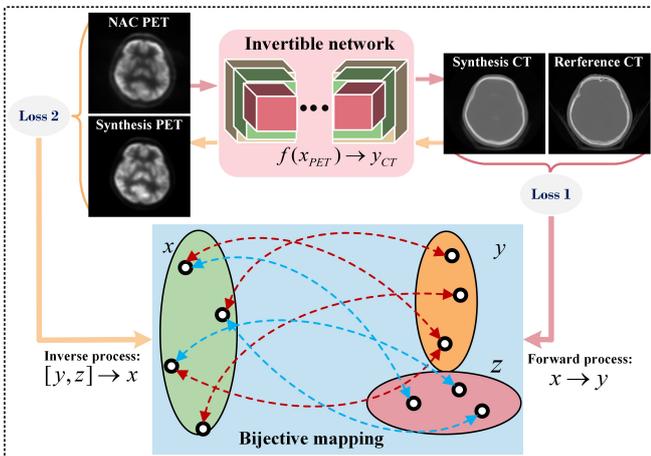

**Fig. 4.** The multi-component loss function of invertible network. Notice that the loss is composed of both forward and inverse process.

## IV. EXPERIMENTAL RESULTS

In this section, the performance of IVNAC is compared with state-of-the-art algorithms (e.g., Cycle-GAN and Pix2pix) on different test images. Comparative experiments verify the effective reconstruction ability of the method. For experiments to be comparable and fair, all methods are conducted on the same dataset.

*A. Experiment Setup*

**Datasets:** The data used in this experiment is a PET/CT image dataset of the brain, which is acquired based on an all-digital PET/CT imaging system (DPET-100 platform provided by RaySolution Digital Medical Imaging Co., Ltd.) with a FOV of $500 \times 500 mm^2$ and an axial FOV of $201.6 mm$. The single-layer PET image matrix is $250 \times 250$, with a pixel size of $2.0 mm \times 2.0 mm$, the single-layer CT image matrix is $512 \times 512$, with a pixel size of $0.9766 mm \times 0.9766 mm$. The data are all obtained from real patients' head PET/CT scans. Due to the uneven response of the PET image system, many edge images obtained are not suitable for training. Therefore, the dataset is built after the edge slices were excluded. A total of 1480 images are collected from 37 patients. Among these patients, 1440 images from 36 patients are selected for the purpose of training, while the remaining 40 images from one patient are designated as a validation dataset for validation of the model during training. Apart from 37 patients, we additionally select a total of 200 images of 5 patients as a test dataset to evaluate the generation performance of the invertible network. Notably, PET images are reconstructed using the ordered-subset expectation maximization (OSEM) algorithm with random correction and normalization, but no attenuation correction is performed. We finely process the images by cropping to preserve the brain information and remove as much tissue outside the head as possible.

**Model Training:** All networks are trained using the Adam solver. We conduct 100 epochs to train the proposed model. The initial learning rate is set to 0.0001 for the first 10 epochs. Every 10 epochs the learning rate is halved. During training, the trade-off parameter $\lambda$ is set to 1. The training and testing experiments are performed with a customized version of Pytorch on an Intel i7-6900K CPU and a GeForce Titan XP GPU.

**Quality Metrics:** It is an open issue to develop objective metrics that correlate with perceived quality measurement. For quality evaluation of synthetic CT images, it should be specific for each application. Following previous state-of-the-art, we perform a quantitative and qualitative assessment of IVNAC. Regarding ground-truth CT as the reference, the peak signal-to-noise ratio (PSNR), structural similarity index (SSIM) and root mean square (RMSE) are employed to evaluate the synthesis performance of the proposed IVNAC algorithm and other methods. For both the metrics, denoting $\tilde{y}$ and $y$ to be the synthesized CT image and the ground-truth, respectively. PSNR is the ratio between the maximum in an image and the intensity of the corrupting noise affecting the fidelity of its representation, it evaluates the noise introduced in the CT synthesis relatively to the ground-truth CT.

$$\text{PSNR}(y, \tilde{y}) = 20 \log_{10} Max(y) / \|\tilde{y} - y\|_2 \qquad (11)$$

SSIM is a more sophisticated metric developed to take advantage of the known characteristics of the human visual system perceiving the loss of image structure due to variations in lighting, i.e.,

$$\text{SSIM}(y, \tilde{y}) = \frac{(2\mu_y \mu_{\tilde{y}} + c_1)(2\sigma_{y\tilde{y}} + c_2)}{(\mu_y^2 + \mu_{\tilde{y}}^2 + c_1)(\sigma_y^2 + \sigma_{\tilde{y}}^2 + c_2)} \qquad (12)$$

RMSE is relatively easy to compute as the average of the absolute difference and difference in quadrature, which is calculated as:

$$\text{RMSE}(y, \tilde{y}) = \sqrt{\sum_1^n (y - \tilde{y})^2 / n} \qquad (13)$$

For PSNR and SSIM, higher values indicate better generation of transformed images. For RMSE, lower values indicate better prediction accuracy for synthetic CT images. Meanwhile, as the end goal of CT synthesis is to use the CT images for PET AC, in order to quantify the performance on PET AC, we also calculated and compared the absolute mean error (MAE) of the brain area in both AC PET and IVNAC PET. $\tilde{v}$ and $v$ indicate IVNAC image and AC PET image, respectively.

$$\text{MAE}(v, \tilde{v}) = \frac{1}{n} \sum_1^n \frac{|\tilde{v} - v|}{v} \qquad (14)$$

*B. Synthetic CT generation and evaluation*

In this subsection, we describe the quantitative and qualitative reconstruction comparisons of IVNAC with two representative methods Cycle-GAN and Pix2pix. The experimental results fully show the comparison of synthetic CT and reference CT of the patients. As shown in Fig. 5, the CT images generated with the proposed method IVNAC share the resemblance in the cranium and brain tissue, which fits the needs of PET AC. Moreover, we further observe from the error map that IVNAC still provides more comparable reconstruction results than Cycle-GAN. Especially on the region of interest and around the edge of the skull, IVNAC reduces most of the errors, while there are relatively higher residual errors in other methods. For example, Pix2pix still has the phenomenon of loss of details. In general, synthetic CT images generated by the proposed method can not only recover bony structures such as cranium, but also demonstrate good contrast on soft tissues around the brain.

TABLE I
GENERATED CT PERFORMANCES ON THE 200 BRAIN CT IMAGES FROM 5 PATIENTS. DATA ARE REPORTED AS MEAN ± STD VALUES.

| Method | PSNR | SSIM | MAE | RMSE |
|---|---|---|---|---|
| IVNAC | **24.00± 1.53** | **0.812± 0.040** | **2.41%± 0.47%** | **6.41%± 1.16%** |
| Cycle-GAN | 18.77± 0.99 | 0.583± 0.028 | 5.62%± 0.77% | 11.60%± 1.40% |
| Pix2pix | 18.59± 1.62 | 0.579± 0.059 | 5.82%± 1.18% | 11.97%± 2.32% |

Output of the validation metrics is listed in Table I. All the quantitative results are the averaged results on the 200 test datasets from 5 patients. Specifically, the average MAE observed on brain synthetic CT images amounts to 2.41%, providing an indication of the overall accuracy of the predictions. Moreover, compared with Cycle-GAN and Pix2pix methods, IVNAC can offer around 6 dB improvement in reconstructed performance in terms of PSNR metrics. It is worth noting that the RMSE exhibits an average value of 6.41%, signifying the performance and precision achieved by the proposed method. The SSIM is close to identity, demonstrating reasonable intensity similarity between the synthetic CT and reference CT. Overall, various quantitative metrics indicate that IVNAC achieves more accurate generation of synthetic CT.

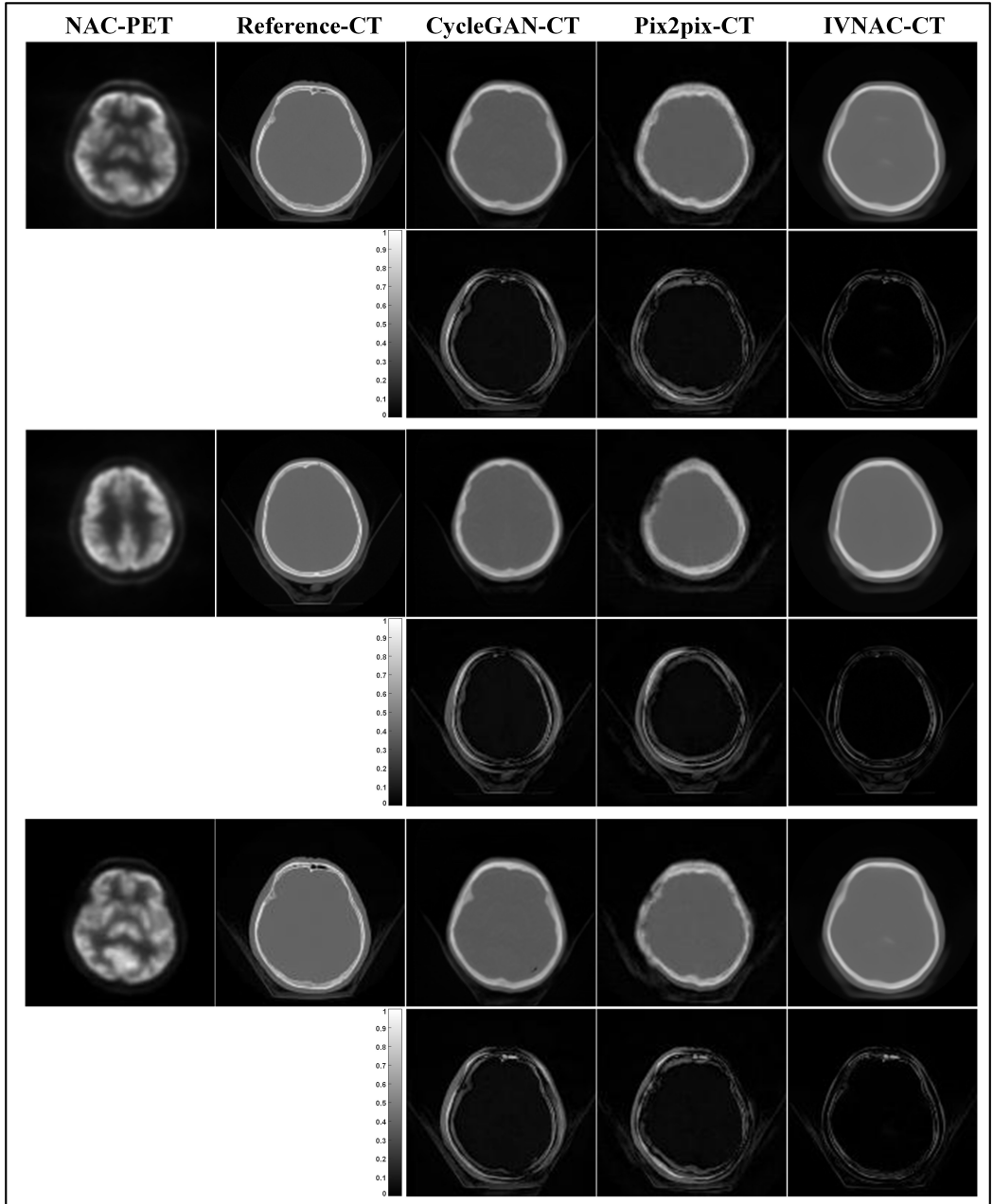

**Fig. 5.** Examples of synthetic CT image on a patient's brain. Five columns from left to right are NAC-PET, reference-CT, Cycle-GAN-CT, Pix2pix-CT and IVNAC-CT, respectively. The second row shows the difference images between the reference CT and the synthetic CT.

*C. Synthetic CT to PET attenuation correction*

The significance of generating synthetic CT images is to perform PET AC, so evaluating the quality of the synthetic CT images depends on the reconstruction results of the NAC PET images. Fig. 6 shows the reconstructed PET image for one patient, which utilizes the synthetic CT and reference CT for PET AC, respectively, as well as pixel-wise relative difference images between these. Apparently, the generated PET images are almost identical to the PET images corrected with reference CT images. From the reconstructed PET images and their corresponding error maps, IVNAC achieves reasonable reconstruction results. Theoretically, the inferiority of Cycle-GAN and Pix2pix indicates that the performance of unsupervised models is not as accurate as that of supervised models in inherent scenarios. Furthermore, excellent agreement is observed in both PET image and error image comparisons. Pix2pix shows blurry edges, which is effectively alleviated by IVNAC.

The fundamental competence of the network lies in its capacity to generate textures and structures that consistent with the original image. Therefore, images with complex details pose a significant challenge to the network's performance. Another challenging case for IVNAC is shown in Fig. 7 for a 48-year-old male with uneven skull edges. From the generated synthetic CT image, it can be seen that although the edge of the restored skull is blurred by the proposed algorithm, the shape and details of the skull are well maintained, so it is very advantageous for PET AC in the later stage. Interestingly, the PET error image also verifies the phenomenon well. Extensive experimental results show that although the generated synthetic brain CT images have a blurring phenomenon at the edge of the skull, there is little effect on PET AC. It means that the PET images after synthetic CT AC demonstrate better matching to the ground truth AC PET images.

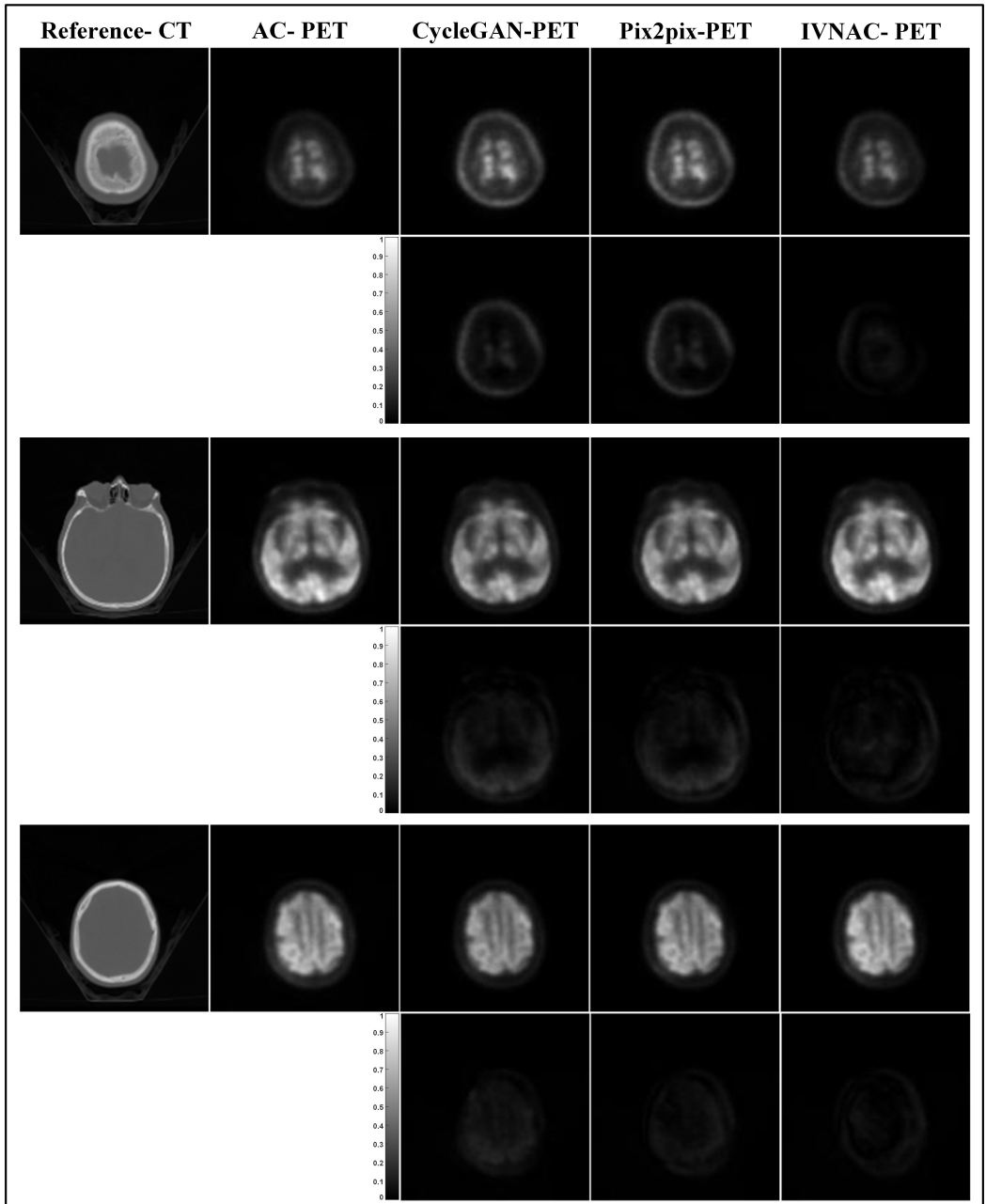

**Fig. 6.** PET data reconstructed with reference and generated synthesized CT images alongside their voxel-wise difference map. Five columns from left to right are reference-CT, AC-PET, Cycle-GAN-PET, Pix2pix-PET and IVNAC-PET, respectively.

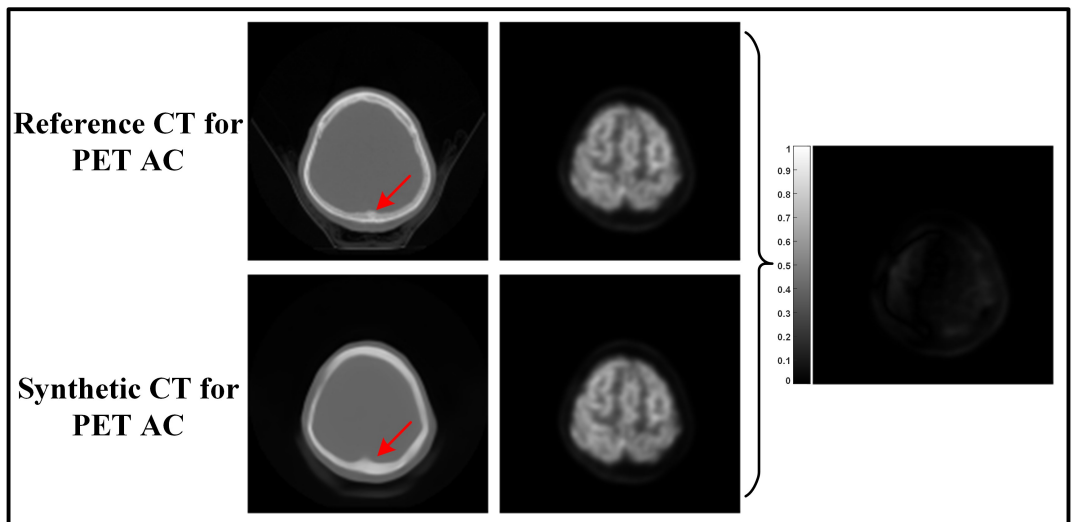

**Fig. 7.** Unique example of PET data reconstructed with reference and generated synthesized CT images alongside their voxel-wise difference map. Red arrows indicate region with relatively obvious details and textures and the intensity of residual maps is two times magnified.

TABLE II
AC PERFORMANCES VIA IVNAC ON 200 BRAIN PET IMAGES FROM 5
PATIENTS. DATA ARE REPORTED AS MEAN ± STD VALUES.

| Method | PSNR | SSIM | MAE | RMSE |
|---|---|---|---|---|
| IVNAC | **38.95± 4.07** | **0.979± 0.009** | **0.35%± 0.15%** | **1.23%± 0.43%** |
| Cycle-GAN | 36.71± 3.02 | 0.978± 0.009 | 0.40%± 0.13% | 1.56%± 0.62% |
| Pix2pix | 36.17± 2.81 | 0.975± 0.009 | 0.43%± 0.13% | 1.64%± 0.59% |

Table II lists the average quantification PET AC results across all scans. As indicated in Table II, the method under investigation exhibits a comparable average PSNR exceeding 38.95 dB, surpassing the comparably lower values of approximately 36.50 dB obtained from Cycle-GAN and Pix2pix. Meanwhile, the SSIM obtained by IVNAC indicates an excellent correlation between the PET images obtained with the proposed method and the reference PET images. Although the quantification on brain PET AC is challenging due to the uneven response of the system, the proposed method is still able to obtain average MAE and RMSE of 0.35% and 1.23%, respectively.

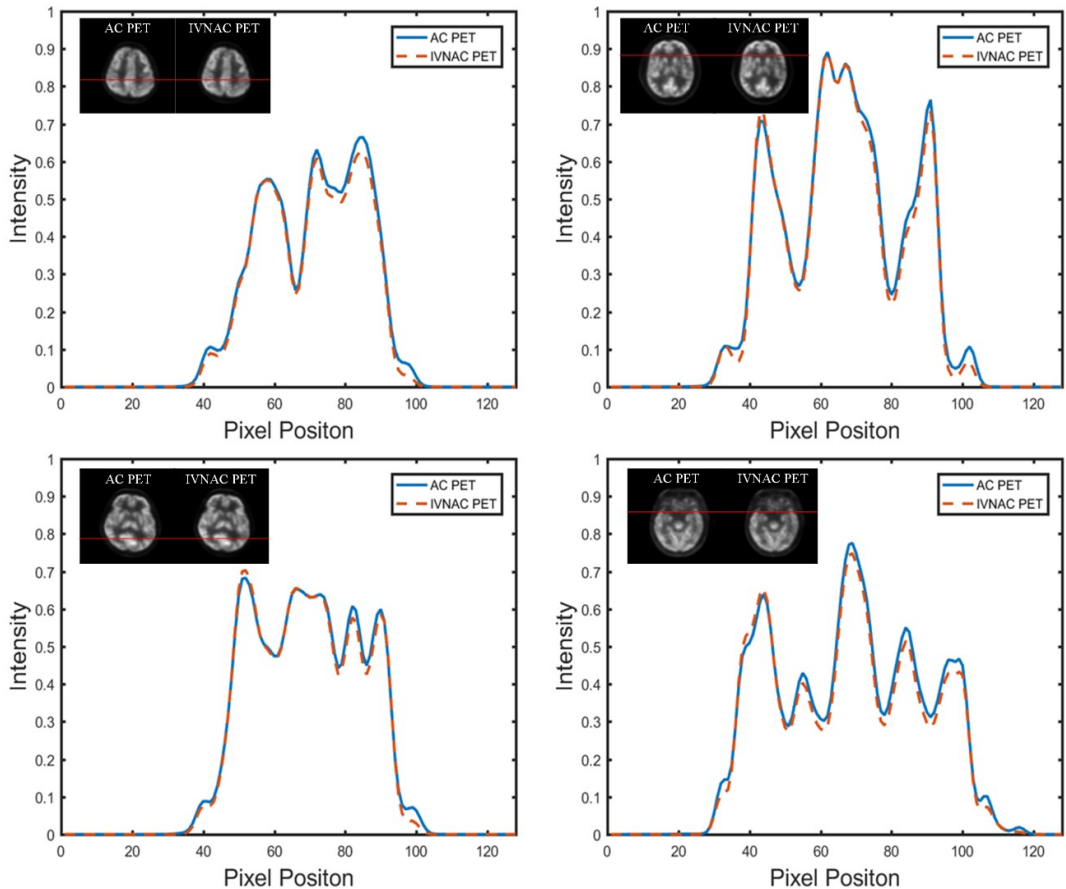

**Fig. 8.** Comparison of PET image profiles between AC-PET and IVNAC-PET on brain images of four different patient. Red lines represent the location where the line profile is measured.

## V. DISCUSSIONS

Our primary contribution is the validation of a brain PET AC method predicated on NAC PET-only image, which provides more accurate synthetic CT images generation by embedding variable augmentation strategy in the framework of invertible network. Generally, the results of the proposed method are competitive compared to other methods in literature. For example, Islam *et al.* proposed a novel approach to generate synthetic CT images using GANs and then created brain PET images for three different stages of Alzheimer's disease [32]. Nevertheless, the quantitative average PSNR of AC PET images obtained by this method was only 32.83dB in the whole image volume, and it can be seen from the visual effect that the details of the AC PET images were relatively blurred. Besides, Dong *et al.* proposed a deep-learning-based approach to create a fully corrected DL-PET dataset from NAC PET by effectively capturing the non-linear relationship between the NAC and AC PET [15].

Moreover, an interesting phenomenon exhibited by this approach would be seen from the PET image profiles that the PET profile generated with synthetic CT AC does not match the reference AC PET profile very well in areas with obvious gray value change. A potential reason can be that the availability of anatomical information on NAC PET images is very limited. Regardless of the challenge, Fig. 8 demonstrates that excellent agreement can be observed from PET image profiles obtained by our method, which indicates great correlation between AC PET and IVNAC PET.

Theoretically, estimating synthetic CT images directly from NAC PET images has encountered much tremendous challenges [13]. Previously, Hu *et al.*, obtained necessary information that was inherently relevant to corresponding CT images after they have already obtained synthetic AC PET images from NAC PET images. Under this condition, although it may theoretically be much easier to estimate the corresponding synthetic CT images from synthetic AC PET images using deep learning methods, quantitative experimental results show that the average PSNR of synthetic CT images obtained by this method was 19.60+1.55dB. In contrast, it can be indicated from the average PSNR value in Table 1 that our algorithm IVNAC can obtain 24.00+1.53dB, which demonstrated that it not only directly generates synthetic CT images from NAC PET images to avoid multiple complex generation issues, but also obtain better results than to estimate synthetic CT from synthetic AC PET directly. In fact, though the similar idea of using only NAC PET to perform AC is proposed previously and has achieved a high level of maturity, improvements to the reconstruction algorithm itself can offer more dramatic improvements. Additionally, the loss curve of the training phase in Fig. 9 indicates a successful training convergence in CT generation where the training loss gradually decreased prior to 60 epochs and reached a plateau after 60 epochs. Hence, it can be seen that IVNAC greatly shortens the generation time and is more stable, which makes a faster generation possible.

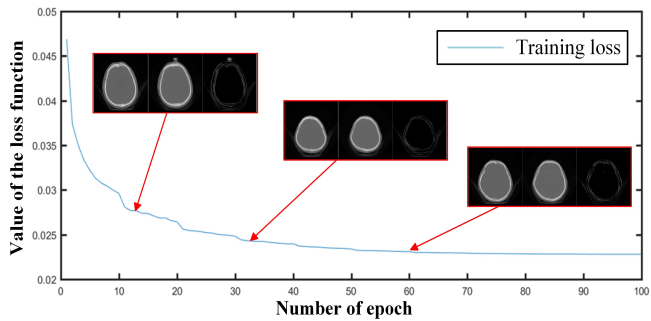

**Fig. 9.** Convergence curves of IVNAC in terms of PSNR versus the iteration number.

Although this work has achieved the abovementioned very good results, there are still some imperfections in our study. In detail, the dataset we used to train the model only contains 40 layers brain PET images of each patient, ignoring the impact of the headrest on the network training. Thence, the currently trained model is not robust enough to provide excellent performance of comprehensive reconstruction. In addition, mismatching problems between reference CT and NAC PET images during training may negatively affect the training and performance of the proposed method. Motivated by the remarkable performances in synthetic CT imaging methods, our future work will focus on CT synthesis with all layers of the brain PET image and other tracers (e.g., DOPA) to improve the robustness of the model.

## VI. CONCLUSIONS

Utilizing deep learning methods to generate synthetic CT images from NAC PET-only images is a promising technique in PET AC. Following the successful completion of network training, the reliance on additional imaging equipment becomes superfluous, thereby effectively mitigating the potential concerns regarding repeated radiation exposure for patients. Moreover, it yields a desirable generation of synthetic CT which ensures satisfactory PET AC results. Therefore, we combine an invertible network with a variable augmentation strategy to directly excavate prior information from NAC PET images to generate synthetic CT images for brain PET AC. Compared to those algorithms that simultaneously obtain both AC PET and attenuation maps from NAC PET images, IVNAC generates synthetic CT images in one step by directly mapping the relationship between two totally different image domains, thus it has more advantages in reducing the computation time. In conclusion, the method IVNAC demonstrates excellent estimation accuracy in brain CT and quantification accuracy in brain PET. Moreover, it exhibits great potential to facilitate AC in brain PET without the need for additional scans.

## ACKNOWLEDGMENTS

The authors sincerely thank the anonymous referees for their valuable comments on this work. They would also like to thank the Institute of Artificial Intelligence, Hefei Comprehensive National Science Center for providing clinical projection data. All authors declare that they have no known conflicts of interest in terms of competing financial interests or personal relationships.